# An Extrasolar Planet Census with a Space-based Microlensing Survey

D.P. Bennett[a], J. Anderson[b], J.-P. Beaulieu[c], I. Bond[d], E. Cheng[e], K. Cook[f], S. Friedman[g], B.S. Gaudi[h], A. Gould[h], J. Jenkins[i], R. Kimble[j], D. Lin[k], M. Rich[l], K. Sahu[g], D. Tenerelli[m], A. Udalski[n], and P. Yock[o]

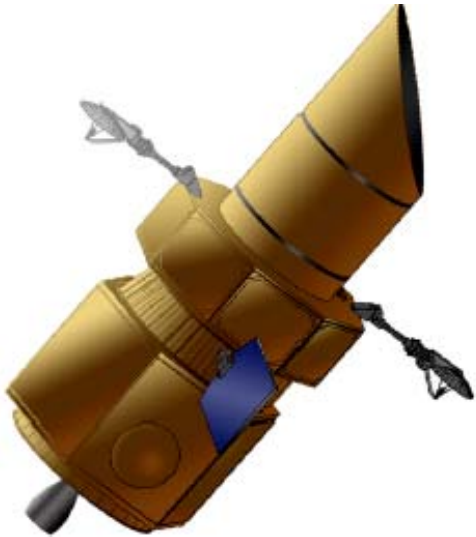
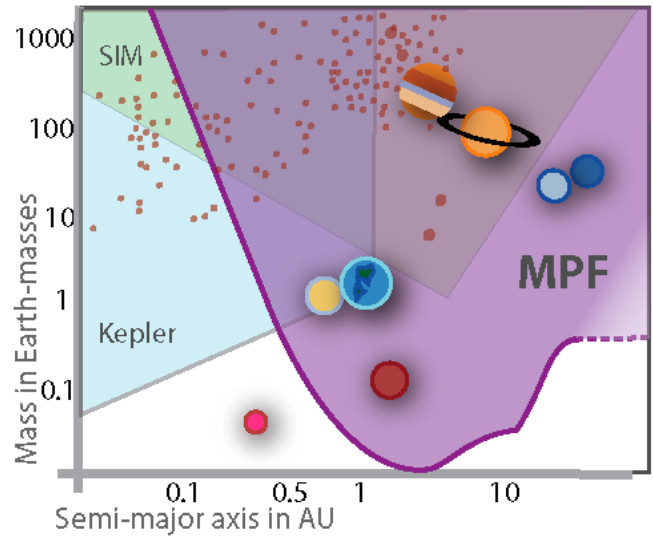

[a] University of Notre Dame, Notre Dame, IN, USA
[b] Rice University, Houston, TX, USA
[c] Institut d'Astrophysique, Paris, France
[d] Massey University, Auckland, New Zealand
[e] Conceptual Analytics, LLC, Glen Dale, MD, USA
[f] Lawrence Livermore National Laboratory, USA
[g] Space Telescope Science Institute, Baltimore, MD, USA
[h] Ohio State University, Columbus, OH, USA
[i] SETI Institute, Mountain View, CA, USA
[j] NASA/Goddard Space Flight Center, Greenbelt, MD, USA
[k] University of California, Santa Cruz, CA, USA
[l] University of California, Los Angeles, CA, USA
[m] Lockheed Martin Space Systems Co., Sunnyvale, CA, USA
[n] Warsaw University, Warsaw, Poland
[o] University of Auckland, Auckland, New Zealand

## ABSTRACT

A space-based gravitational microlensing exoplanet survey will provide a statistical census of exoplanets with masses $\geq 0.1 M_\oplus$ and orbital separations ranging from 0.5AU to $\infty$. This includes analogs to all the Solar System's planets except for Mercury, as well as most types of planets predicted by planet formation theories. Such a survey will provide results on the frequency of planets around all types of stars except those with short lifetimes. Close-in planets with separations < 0.5 AU are invisible to a space-based microlensing survey, but these can be found by Kepler. Other methods, including ground-based microlensing, cannot approach the comprehensive statistics on the mass and semi-major axis distribution of extrasolar planets that a space-based microlensing survey will provide. The terrestrial planet sensitivity of a ground-based microlensing survey is limited to the vicinity of the Einstein radius at 2-3 AU, and space-based imaging is needed to identify and determine the mass of the planetary host stars for the vast majority of planets discovered by microlensing. Thus, a space-based microlensing survey is likely to be the only way to gain a comprehensive understanding of the nature of planetary systems, which is needed to understand planet formation and habitability. The proposed Microlensing Planet Finder (MPF) mission is an example of a space-based microlensing survey that can accomplish these objectives with proven technology and a cost that fits comfortably under the NASA Discovery Program cost cap.

## 1. Basics of the Gravitational Microlensing Method

The physical basis of microlensing is the gravitational attraction of light rays by a star or planet. As illustrated in Fig. 1, if a "lens star" passes close to the line of sight to a more distant source star, the gravitational field of the lens star will deflect the light rays from the source star. The gravitational bending effect of the lens star "splits", distorts, and magnifies the images of the source star. For Galactic microlensing, the image separation is $\leq 4$ mas, so the observer sees a microlensing event as a transient brightening of the source as the lens star's proper motion moves it across the line of sight.

Gravitational microlensing events are characterized by the Einstein ring radius,

$$R_E = 2.0 \text{ AU} \sqrt{\frac{M_L}{0.5 M_\odot} \frac{D_L(D_S - D_L)}{D_L(1 \text{ kpc})}},$$

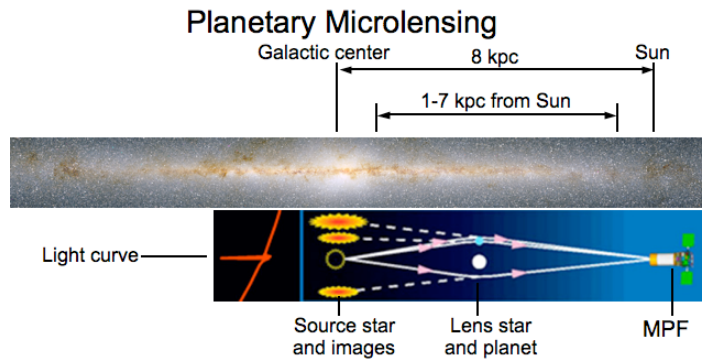

**Fig. 1:** *The geometry of a microlensing planet search towards the Galactic bulge. Main sequence stars in the bulge are monitored for magnification due to gravitational lensing by foreground stars and planets in the Galactic disk and bulge.*

where $M_L$ is the lens star mass, and $D_L$ and $D_S$ are the distances to the lens and source, respectively. This is the radius of the ring image that is seen with perfect alignment between the lens and source stars. The lensing magnification is determined by the alignment of the lens and source stars measured in units of $R_E$, so even low-mass lenses can give rise to high magnification microlensing events. The duration of a microlensing event is given by the Einstein ring crossing time, which is typically 1-3 months for stellar lenses and a few days or less for a planet.

**Planets are detected via light curve deviations** that differ from the normal stellar lens light curves (Mao & Paczynski 1991). Usually, the signal occurs when one of the two images due to lensing by the host star passes close to the location of the planet, as indicated in Fig. 1 (Gould & Loeb 1992), but planets can also be detected at very high magnification where the gravitational field of the planet destroys the symmetry of the Einstein ring (Griest & Safizadeh 1998).

## 2. Capabilities of the Microlensing Method

**Planets down to one tenth of an Earth mass can be detected**. The probability of a detectable planetary signal and its duration both scale as $R_E \sim M_p^{1/2}$, but given the optimum alignment, planetary signals from low-mass planets can be quite strong. The limiting mass for the microlensing method occurs when the planetary Einstein radius becomes smaller than the projected radius of the source star (Bennett & Rhie 1996). The ~5.5 $M_\oplus$ planet detected by Beaulieu et al. (2006) is near this limit for a giant source star, but most microlensing events have G or K-dwarf source stars with radii that are at least 10 times smaller than this. So, the sensitivity of the microlensing method extends down to < 0.1$M_\oplus$, as the results of a detailed simulation of the MPF mission (Bennett & Rhie 2002) show in Fig. 2.

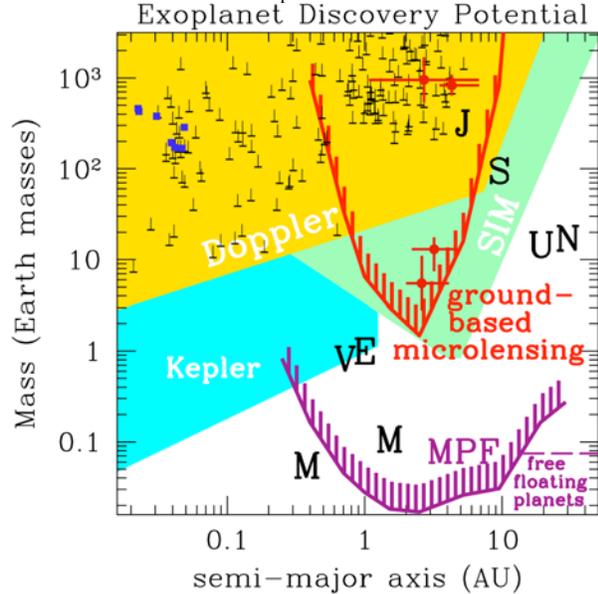

**Fig. 2:** *MPF is sensitive to planets above the purple curve in the mass vs. semi-major axis plane. The gold, green and cyan regions indicates the sensitivities of radial velocity surveys, SIM and Kepler, respectively. The location of our Solar System's planets and many extrasolar planets are indicated, with ground-based microlensing discoveries in red.*

**Microlensing is sensitive to a wide range of planet-star separations and host star types.** The host stars for planets detected by microlensing are a random sample of stars that happen to pass close to the line-of-sight to the source stars in the Galactic bulge, so all common types of stars are surveyed, including G, K, and M-dwarfs, as well as white dwarfs and brown dwarfs. Microlensing is most sensitive to planets at a separation of ~$R_E$ (usually 2-3 AU) due to the strong stellar lens magnification at this separation, but the sensitivity extends to arbitrarily large separations. It is only planets well inside $R_E$ that are missed because the stellar lens images that would be distorted by these inner planets have very low magnifications and a very small contribution to the total brightness. These features can be seen in Fig. 2, which compares the sensitivity of the MPF mission with expectations for other planned and current programs. Other ongoing and planned programs can detect, at most, analogs of two of the Solar System's planets, while a space-based microlensing survey can detect seven—all but Mercury. The only method with comparable sensitivity to MPF is the Kepler space-based transit survey, which complements the microlensing method with sensitivity at semi-major axes, $a \leq 1$. The sensitivities of MPF and Kepler overlap at separations of ~1 AU, which corresponds to the habitable zone for G and K stars.

The red crosses in Fig. 2 indicate the two gas giant (Bond et al. 2004; Udalski et al. 2005) and two ~$10M_\oplus$ "super-earth" planets in orbits of ~3 AU discovered by ground-based microlensing (Beaulieu et al. 2006; Gould et al. 2006). A preliminary analysis suggests that about one third of all stars are likely to have a super-earth at 1.5-4AU whereas radial velocity surveys find that only about 3% of stars have gas giants in this region (Butler et al. 2006).

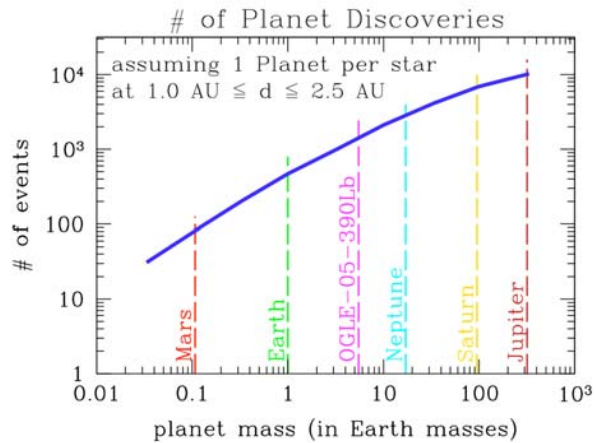

*Fig. 3:* The expected number of MPF planet discoveries as a function of the planet mass if every star has a single planet at a separation of 1.0-2.5 AU.

**Microlensing light curves yield unambiguous planet parameters.** For the great majority of events, the basic planet parameters (planet:star mass ratio, planet-star separation) can be "read off" the planetary deviation (Gould & Loeb 1992; Bennett & Rhie 1996; Wambsganss 1997). Possible ambiguities in the interpretation of planetary microlensing events have been studied in detail (Gaudi & Gould 1997; Gaudi 1998), and these can be resolved with good quality, continuous light curves that will be routinely acquired with a space-based microlensing survey. A space-based survey will also detect most of the planetary host stars, which generally allows the host star mass, approximate spectral type, and the planetary mass and separation to be determined (Bennett et al. 2007). The distance to the planetary system is determined when the host star is identified, so a space-based microlensing survey will also measure how the properties of exoplanet systems change as a function of distance from the Galactic Center. There is usually some redundancy in the measurements that determine the properties of the host stars, and so the determination is robust to complicating factors, such as a binary companion to the background source star.

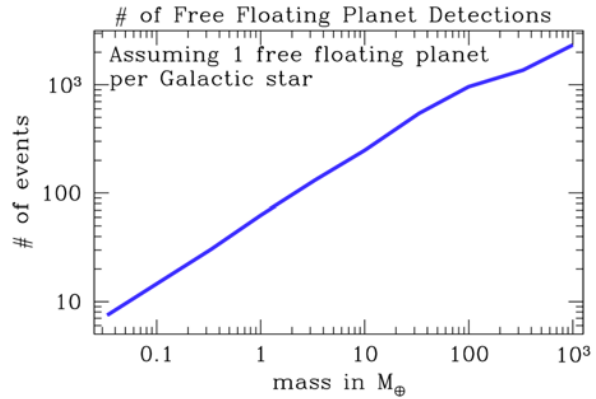

*Fig. 4:* The expected number of MPF free-floating planet discoveries.

**Detailed simulations indicate a large number of planet detections.** Bennett & Rhie (2002) and Gaudi (unpublished) have independently simulated space-based microlensing surveys. These simulations included variations in the assumed mission capabilities that allow us to explore how changes in the mission design will affect the scientific output, and they form the basis of our predictions in Figs. 2-4. In order to predict the number of planets that will be detected by a space-based microlensing survey, we must make assumptions regarding the frequency of exoplanets. Figs. 3 and 4 show the expected number of planets that MPF would detect at orbital separations 1-2.5 AU and ∞ (i.e. free-floating planets) assuming one such planet per star. The range 1-2.5 AU is presented because this is just outside the range of Kepler and inside the region of highest sensitivity for ground-based microlensing surveys. It also corresponds to the outer part of the habitable zone for G and K stars and contains the "snow-line" for part of the history of lower mass stars (Kennedy et al. 2006). Free-floating planets are expected to be a common by-product of most planet formation scenarios (Levison et al. 1998; Goldreich et al. 2004), and only a space-based microlensing survey can detect free-floating planets of $\leq 1 M_\oplus$.

## 3. A Space-based Microlensing Survey Is Needed

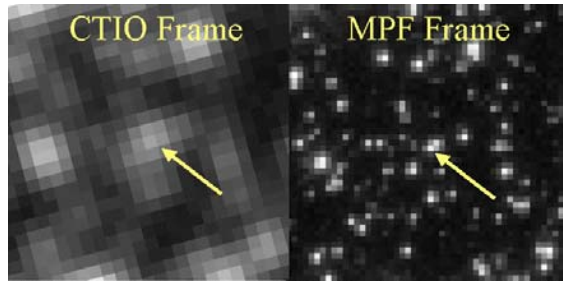

Fig. 5: A comparison between an image of the same star field in the Galactic bulge from CTIO in 1" seeing and a simulated MPF frame (based on an HST image). The indicated star is a microlensed main sequence source star.

Microlensing relies upon the high density of source and lens stars towards the Galactic bulge to generate the stellar alignments that are needed to generate microlensing events, but this high star density also means that the bulge main sequence source stars are not generally resolved in ground-based images, as Fig. 5 demonstrates. This means that the precise photometry needed to detect planets of $\leq 1 M_\oplus$ is not possible from the ground unless the magnification due to the stellar lens is moderately high. This, in turn, implies that ground-based microlensing is only sensitive to terrestrial planets located close to the Einstein ring (at ~2-3 AU). The full sensitivity to terrestrial planets in all orbits from 0.5 AU to ∞ comes only from a space-based survey.

**Planetary host star detection from space yields precise star and planet parameters.** For all but a small fraction of planetary microlensing events, space-based imaging is needed to detect the planetary host stars, and the detection of the host stars allows the star and planet masses and separation in physical units to be determined. This can be accomplished with HST observations for a small number of planetary microlensing events (Bennett et al. 2006), but space-based survey data will be needed for the detection of host stars for hundreds or thousands of planetary microlensing events. Fig. 6 shows the distribution of planetary host star masses and the predicted uncertainties in the masses and separation of the planets and their host stars (Bennett et al. 2007) from simulations of the MPF mission. The host stars with masses determined to better than 20% are indicated by the red histogram in Fig. 6(a), and these are primarily the host stars that can be detected in MPF images.

Ground-based microlensing surveys also suffer significant losses in data coverage and quality due to poor weather and seeing. As a result, a significant fraction of the planetary deviations seen in a ground-based microlensing survey will have poorly constrained planet parameters due to poor light curve coverage (Peale 2003).

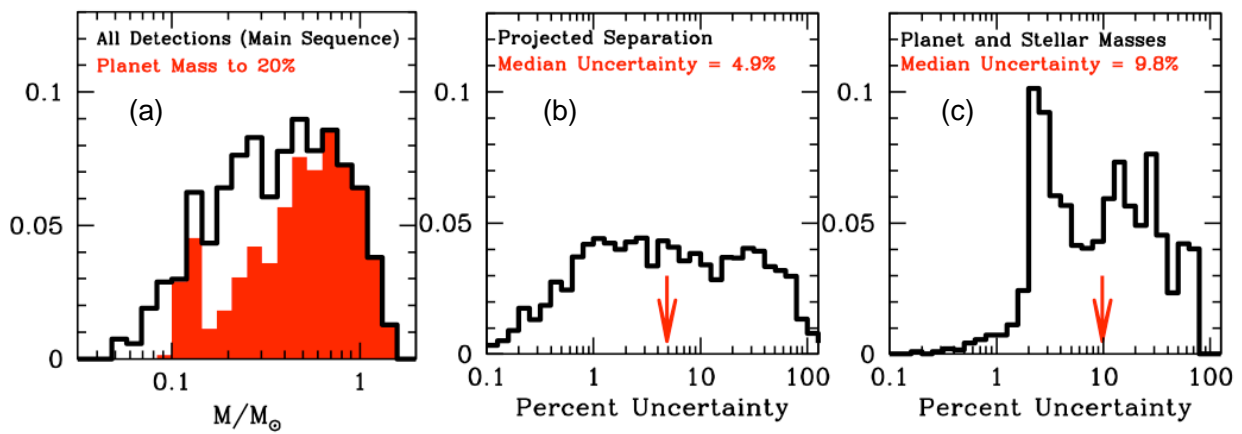

*Fig. 6:* (a) The simulated distribution of stellar masses for stars with detected terrestrial planets. The red histogram indicates the subset of this distribution for which the masses can be determined to better than 20%. (b) The distribution of uncertainties in the projected star-planet separation. (c) The distribution of uncertainties in the star and planet masses.

## 4. A Space-Based Microlensing Survey Constrains Planet Formation Theories

Rapid advancement in exoplanet research is driven by both extensive observational searches around mature stars as well as the construction of planet formation and evolution models. Perhaps the most surprising discovery so far is the great diversity in the planets' dynamical properties, but these results are largely confined to planets that are unusually massive or reside in very close orbits. The core accretion theory suggests most planets are much less massive than gas giants and that the critical region for understanding planet formation is the "snow-line", located in the region (1.5-4 AU) of greatest microlensing sensitivity (Ida & Lin 2005; Kennedy et al. 2006). Early results from ground-based microlensing searches (Beaulieu et al. 2006; Gould et al. 2006) appear to confirm these expectations. A space-based microlensing survey would extend the current sensitivity of the microlensing method down to masses of ~$0.1M_\oplus$ over a large range (0.5AU-∞) in separation, and in combination with Kepler, such a mission provides sensitivity to sub-Earth mass planets at all separations. The semi-major axis region probed by space-microlensing provides a cleaner test of planet formation theories than the close-in planets detected by other methods, because planets discovered at > 0.5 AU are more likely to have formed *in situ* than the close-in planets. The sensitivity region for space-microlensing includes the outer habitable zone for G and K stars through the "snow-line" and beyond, and the lower sensitivity limit reaches the regime of planetary embryos at ~$0.1M_\oplus$. It may be that such planets are much more common than planets of $1M_\oplus$ because their type-1 migration time is much longer.

**Space-microlensing tests core accretion.** The space-microlensing census of low-mass planets should also provide direct evidence of features of the proto-planetary disk predicted by the core accretion theory. There are several physical processes that control the development of planetary embryos and planets in the proto-planetary disk. In the inner disk, the size of planetary embryos is controlled by the isolation mass, and the isolation mass is expected to jump by an order of magnitude across the "snow-line" because of the increased surface density of solids in the disk. But the number of gas giant and super-earth planets is also expected to increase beyond the snow line, while the planetary growth time increases. This means that it is more likely for the growth of outer planets to be terminated via gravitational scattering of planetesimals or the proto-planets themselves. Scattering would also result in the removal of lower mass planets into very distant orbits or even out of the gravitational influence of the host stars altogether, but space-based microlensing can still detect planets in these locations. The frequency of planets of different masses and separations that a space-based microlensing survey provides will yield a unique insight into the planetary formation process and will allow us to determine the importance of these processes.

**The habitability of a planet depends on its formation history.** The suitability of a planet for life depends on a number of factors, such as the average surface temperature, which determines if the planet resides in the habitable zone. However, there are many other factors that also may be important, such as the presence of sufficient water and other volatile compounds necessary for life (Raymond et al. 2004; Lissauer 2007). Thus, a reasonable understanding of planet formation is an important foundation for the search for nearby habitable planets and life.

## 5. Overview of the Microlensing Planet Finder Mission

Key requirements of the MPF mission are summarized in table 1. MPF continuously observes four 0.65 sq. deg. fields in the central Galactic Bulge using an inclined geostationary orbit to provide a continuous view of the Galactic Bulge fields and a continuous downlink. MPF will use a dedicated ground station co-located with other NASA facilities at White Sands, NM. Spacecraft commanding and on-board processing are minimized because of the simple observation plan and orbit design.

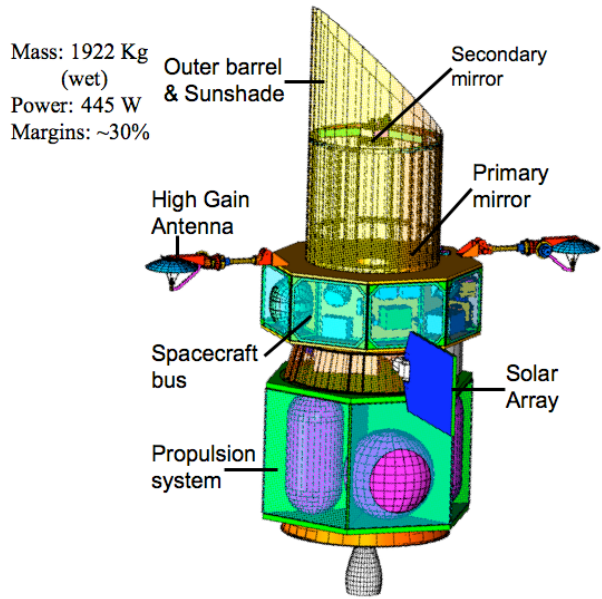

*Fig. 7: MPF On-Orbit Configuration*

| Property | Value | Units |
|---|---|---|
| Launch Vehicle | 7920-9.5 | Delta II |
| Orbit | Inclined GEO 28.7 | degrees |
| Mission Lifetime | 4.0 | years |
| Telescope Aperture | 1.1 | meters (diam.) |
| Field of View | 0.95x0.68 | degrees |
| Spatial Resolution | 0.240 | arcsec/pixel |
| Pointing Stability | 0.048 | arcsec |
| Focal Plane Format | 145 | Megapixels |
| Spectral Range | 600 – 1700 | nm in 3 bands |
| Quantum Efficiency | > 75% | 900-1400 nm |
|  | > 55% | 700-1600 nm |
| Dark Current | < 1 | e-/pixel/sec |
| Readout Noise | < 30 | e-/read |
| Photometric Accuracy | 1 or better | % at J=20.5. |
| Data Rate | 50.1 | Mbits/sec |

*Table 1: Key MPF Mission Requirements*

**MPF system.** MPF uses a 1.1m Three-Mirror Anastigmat (TMA) telescope feeding a 145 Mpixel HgCdTe focal plane residing on a standard spacecraft bus as shown in Fig. 7. The MPF design leverages existing hardware and design concepts, many of which are already demonstrated on-orbit and/or flight qualified. The spacecraft bus is a near-identical copy of that used for Spitzer and has demonstrated performance that meets MPF requirements. The telescope system leverages Ikonos and NextView commercial Earth-observing telescope designs that provide extensive diffraction-limited images. The focal plane design taps proven technologies developed for JWST. All elements are at TRL 6 or better. The focal plane design uses common non-destructive readout CMOS multiplexers for two detector technologies that cover the visible and the near-IR. The MPF focal plane can track up to 35 guide stars in the field, providing a built-in fine guidance capability. The focal plane gains additional advantages from using the Teledyne SIDECAR™ application specific integrated circuit (ASIC) that condenses all the control and readout electronics into a "system-on-a-chip" implementation. This approach dramatically simplifies the support electronics while minimizing wire-count challenges.

**Cost and Schedule.** The total cost for the MPF mission is (FY06) $390M including 30% contingency during development. The team that developed the costs included NASA GSFC, Lockheed Martin, ITT, STScI, Teledyne and University of Notre Dame.

## 6. Discussion and Summary

A space-based microlensing survey provides a census of extrasolar planets that is complete (in a statistical sense) down to $0.1 M_\oplus$ at orbital separations $\geq 0.5$ AU, and when combined with the results of the Kepler mission a space-based microlensing survey will give a comprehensive picture of all types of extrasolar planets with masses down to well below an Earth mass. This complete coverage of planets at all separations can be used to calibrate the poorly understood theory of planetary migration. This fundamental exoplanet census data is needed to gain a comprehensive understanding of processes of planet formation and migration, and this understanding of planet formation is an important ingredient for the understanding of the requirements for habitable planets and the development of life on extrasolar planets.

A subset of the science goals can be accomplished with an enhanced ground-based microlensing program (Gould et al. 2007), which would be sensitive to Earth-mass planets in the vicinity of the "snow-line". But such a survey would have its sensitivity to Earth-like planets limited to a narrow range of semi-major axes, so it would not provide the complete picture of the frequency of exoplanets down to $0.1 M_\oplus$ that a space-based microlensing survey would provide. Furthermore, a ground-based survey would not be able to detect the planetary host stars for most of the events, and so it will not provide the systematic data on the variation of exoplanet properties as a function of host star type that a space-based survey will provide.

The basic requirements for a space-based microlensing survey are a 1-m class wide field-of-view space telescope that can image the central Galactic bulge in the near-IR or optical almost continuously for periods of at least several months at a time. This can be accomplished as a NASA Discovery mission, as the example of the MPF mission shows, but there are a number of other proposed missions with similar requirements, such as a number of JDEM concept missions or a stare-mode astrometry mission (Johnston et al. 2007) that makes use of the same type of detectors as the MPF mission. Such an astrometry mission could complement the statistical planetary results from microlensing survey with data on nearby planets. Thus, a space-based microlensing survey could be accomplished with a standalone Discovery class mission or a joint mission with another project. As Fig. 2 shows, there is no other planned mission that can duplicate the science return of a space-based microlensing survey, and our knowledge of exoplanets and their formation will remain incomplete until such a mission is flown.